\documentclass[aps,11pt,onecolumn,preprintnumbers,amsmath,amssymb]{revtex4}

\usepackage[english]{babel}
\usepackage{graphicx}
\usepackage{color}

\begin{document}

\newcommand{\unit}[1]{\:\mathrm{#1}}            
\newcommand{\To}{\mathrm{T_0}}
\newcommand{\Tp}{\mathrm{T_+}}
\newcommand{\Tm}{\mathrm{T_-}}
\newcommand{\EST}{E_{\mathrm{ST}}}
\newcommand{\Rp}{\mathrm{R_{+}}}
\newcommand{\Rm}{\mathrm{R_{-}}}
\newcommand{\Rpp}{\mathrm{R_{++}}}
\newcommand{\Rmm}{\mathrm{R_{--}}}
\newcommand{\ddensity}[2]{\rho_{#1\,#2,#1\,#2}} 
\newcommand{\ket}[1]{\left| #1 \right>} 
\newcommand{\bra}[1]{\left< #1 \right|} 

\title{Quantum Teleportation from a Propagating Photon to a Solid-State Spin Qubit}
\author{W. B. Gao}
\author{P. Fallahi}
\author{E. Togan}
\author{A. Delteil}
\author{Y. S. Chin}
\author{J. Miguel-Sanchez}
\author{A. Imamo\u{g}lu}
\affiliation{Institute of Quantum Electronics, ETH Zurich, CH-8093
Zurich, Switzerland.}


\maketitle

\textbf{The realization of a quantum interface between a propagating
photon used for transmission of quantum information, and a
stationary qubit used for storage and manipulation, has long been an
outstanding goal in quantum information science \cite{Divincenzo00}.
A method for implementing such an interface between dissimilar
qubits is quantum teleportation \cite{Bouwmeester97,Sherson06,Yuao08}, which has
attracted considerable interest not only as a versatile
quantum-state-transfer method but also as a quantum computational
primitive \cite{Gottesman99,KLM01}. Here, we experimentally demonstrate
transfer of quantum information carried by a photonic qubit to a
quantum dot spin qubit using quantum teleportation. In
our experiment, a single photon in a superposition state of two
colors -- a photonic qubit -- is generated using selective resonant
excitation of a neutral quantum dot. We achieve an unprecedented
degree of indistinguishability of single photons from different
quantum dots by using local electric and magnetic field control. To
teleport a photonic qubit, we generate an entangled spin-photon
state \cite{Gao12,DeGreve12,Schaibley13} in a second quantum dot
located 5 meters away from the first and interfere the photons from
the two dots in a Hong-Ou-Mandel set-up. A coincidence detection at
the output of the interferometer heralds successful teleportation,
which we verify by measuring the resulting spin state after its
coherence time is prolonged by an optical spin-echo pulse sequence.
The demonstration of successful inter-conversion of photonic and
semiconductor spin qubits constitute a major step towards the
realization of on-chip quantum networks based on semiconductor
nano-structures \cite{Cirac97}.}

Semiconductor quantum dots (QD) have been at the forefront of two
seemingly uncorrelated approaches to experimental quantum
information science \cite{Divincenzo00}: on the one hand, single QDs
in photonic nanostructures have been used to realize high-efficiency
sources for single-photons \cite{Michler00,Santori02,Claudon10} with
potential applications in quantum key distribution \cite{BB84} and
linear optics quantum computation \cite{KLM01}. On the other hand,
single spins confined in quantum dots have been extensively studied
as promising solid-state qubits with long coherence time and a
potential for integration on a chip
\cite{Loss98,Imamoglu99,Kroutvar04,Petta05,Press08,Greilich09}. By
realizing teleportation of the quantum state of a single photon in a
superposition of two frequency components generated by one QD to the
spin-state of another QD located in a different cryostat, we
demonstrate that these two auspicious features of QDs can be
combined to realize an elementary process relevant for a quantum
network \cite{Cirac97}.

The key step in photon-to-spin quantum teleportation protocol
\cite{Bouwmeester97} is the  two-photon interference of the photon
(in mode A) to be teleported with a second photon (in mode B) that
is initially in a maximally entangled state with the target spin
state (Fig.~1). Detection of a coincidence event at the output of a
Hong-Ou-Mandel (HOM) interferometer heralds the successful
teleportation of the quantum state of the single photon in mode A,
given by the arbitrary superposition of its two frequency
components, $|\psi_p> = \alpha |\omega_b>_A + \beta |\omega_r>_A$,
to the QD electron spin-state, which after the completion of the
protocol reads $|\psi_e> = \alpha |\downarrow> + \beta |\uparrow>$.

To generate a photonic qubit, we use a neutral self-assembled InGaAs
QD (QD1 or QD3) where the elementary optical excitations from the
unique ground state $|0>$ are the two fundamental exciton states
$X_r$ and $X_b$ that are split in energy by the anisotropic
electron-hole exchange interaction. The exciton state $X_r$ ($X_b$)
decays at rate $\Gamma_1$ by spontaneous emission of a photon at
frequency $\omega_r$ ($\omega_b$) back into $|0>$ (Fig.~2a). A laser
field with a pulse duration short compared to $\Gamma_1^{-1}$ and
 resonant with either of these two exciton states results in
the generation of a single-color single photon pulse. Alternatively,
applying a two-color laser pulse that is simultaneously resonant
with both $X_r$ and $X_b$ results in the generation of a
single-photon that is in a superposition of the two central
frequency components $\omega_r$ and $\omega_b$. The left-hand side
of Figure~1 shows the experimental set-up we use to generate such a
single-photon qubit. An amplitude electro-optic modulator (EOM) is
used to generate $400$ps pulses from a continuous-wave (cw) laser
field. A second EOM is used to generate sidebands that enable us to
resonantly drive $X_r$ and $X_b$ simultaneously (Supplementary Information). The
generated pulse is depicted in Figure~2a: the beats at
$\omega_b-\omega_r$ demonstrate that the single-photon is in a
coherent superposition of two frequencies. Photon correlation
measurements carried out on the emitted light in turn show strong
antibunching, proving that our scheme generates a nearly ideal
single-photon pulse (Fig.~2a inset).

A high teleportation fidelity requires a high degree of
indistinguishability of the two photons  generated by two QDs placed
in separate cryostats. To  determine the degree of
indistinguishability we attain, we use single photons generated by
two neutral QDs, QD1 and QD2, each tuned using external electric and
magnetic fields to ensure that their transition frequencies
$\omega_r$ and $\omega_b$ are identical. Figure~2b shows the time
resolved coincidences between the two output ports of the HOM
interferometer as a function of the time delay between the photon
detection times, when the two photonic qubits generated with either
identical or orthogonal polarizations are sent to the input ports of
the interferometer. Our measurements reveal that the visibility is
$V=(C_\bot-C_\|)/C_\bot= 80.2\pm 2.9 \%$, where $C_{\|}$
($C_{\bot}$) denotes the integrated number of counts in [-1.2 ns,
1.2 ns] time window when the photon polarizations in the input ports
are identical (orthogonal). This represents the highest value
reported to date for photons generated by two different QDs (see
Sec.~C of the Supplementary Information) \cite{Flagg10,Patel10}.
Introducing a half period time-delay for one of the single-photon
pulses also renders the photonic qubits distinguishable (Fig.~2b).

To generate an entangled spin-photon pair, we use the trion emission
from the  single-electron charged state of QD2
\cite{Gao12,DeGreve12,Schaibley13}, driven using a combination of
resonant excitation and non-resonant rotation pulses as depicted in
the right hand side of Fig.~1. The photonic qubit that is coupled to
the mode A is generated by the neutral QD3, which is in the same
cryostat as QD1 and whose exciton transition energy is nearly
identical to that of the QD2 trion.
Figure~2c shows the relevant energy-level diagram as well as the
allowed optical transitions for QD2 in Voigt geometry where an
external magnetic field $B_x= 0.7$T is applied perpendicular to the
growth direction. To ensure that $\omega_b-\omega_r$ of QD3 matches
the electron Zeeman energy of QD2, a separate magnetic field of 0.12
T is applied to QD3. Moreover, the electric fields applied
separately to the QD2 and QD3 are adjusted such that their emission
frequencies ($\omega_b$ and $\omega_r$) are identical.  The ground
states of the QD are identified by the orientation of the electron
spin, with $| \uparrow \rangle$ ($|\downarrow \rangle$) denoting
spin parallel (anti-parallel) to the magnetic field direction.
Spontaneous emission of H (V) polarized photon at frequency
$\omega_{r}$ ($\omega_{b}$) from the trion state $| T_r\rangle$ at
rate $\Gamma_2 / 2$ brings the QD back into the $| \downarrow
\rangle$ ($|\uparrow \rangle$) state, resulting in the maximally
entangled state \cite{Gao12}
\begin{equation}
|\Psi\rangle = \frac{1}{\sqrt{2}} ( |\downarrow\rangle |\omega_{r}; H\rangle_B +i|\uparrow\rangle |\omega_{b} ;
V\rangle_B ) \;\;\;. \label{eq1}
\end{equation}
After erasing the polarization information using a  polarizer
transmitting $H-iV$ polarized light, the spin-photon entangled state
reads $|\Psi\rangle = \frac{1}{\sqrt{2}} ( |\downarrow\rangle
|\omega_{r}\rangle_B -|\uparrow\rangle |\omega_{b}\rangle_B )$
\cite{Gao12}. Here, we indicated that the photon that is entangled
with the spin is channeled into the photonic mode B, which
constitutes the second input mode of the HOM interferometer
(Fig.~1).

Hyperfine interaction between the electron spin and the QD nuclei
leads to a spin-decoherence time of $T_2^* \sim 1$ns. To ensure that
the electron spin coherence is intact when the coincidence detection
in the HOM interferometer heralds successful teleportation in $\sim
11$ns after the generation of the entangled spin-photon state, we
introduce a spin-echo scheme that extends the coherence of the
electron spin \cite{Press10}. For an echo time $T_{echo} = 13$ns,
measurement of the spin-photon correlations in the rotated basis
demonstrates that spin-photon entanglement fidelity is $F > 0.63 \pm
0.02$ (Fig.~2c); this lower bound for the fidelity is determined
predominantly by the detection jitter ($60ps$)~ (Supplementary Information). We
emphasize that despite the similarity with the single-photon
pulse-shape in Fig.~2a, what is depicted in Fig.~2c is the time
dependence of the coincidence counts as a function of generation
time of the entangled spin-photon pair, with $t=0$ denoting the
rise-time of the entanglement generation pulse (see the pulse
sequence depicted in Fig.~2c).


The state of the coupled system consisting of two photons in modes A
$\&$ B and the QD spin prior to the beam splitter is
\begin{equation}
  |\Psi\rangle=(\alpha|\omega_{b}\rangle_{A}+\beta|\omega_{r}\rangle_{A})\otimes \frac{1}{\sqrt{2}}(|\omega_{r}\rangle_{B}|\downarrow\rangle-|\omega_{b}\rangle_{B}|\uparrow\rangle)
  \label{eq2}
\end{equation}
If the photons in modes A $\&$ B are indistinguishable in every
aspect but their internal (color) state, the only possibility for a
simultaneous coincidence detection at the output of the HOM
interferometer is to have the input two-photon state in
$|\varphi_S\rangle=\frac{1}{\sqrt{2}}(|\omega_{b}\rangle_{A}|\omega_{r}\rangle_{B}-|\omega_{r}\rangle_{A}|\omega_{b}\rangle_{B})$
\cite{Olmschenk09}. Therefore, detection of a coincidence projects
the input photonic state (in modes A$\&$B) to $|\varphi_S\rangle$
\cite{Bouwmeester97}. The  spin-state corresponding to this
measurement outcome is
$\langle\varphi_S|\Psi\rangle=\alpha|\uparrow\rangle +
\beta|\downarrow\rangle$, as can be verified from Eq.~\ref{eq2}.
Experimental verification of teleportation is based on three-fold
coincidence detection of photons at the the two output modes of the
HOM interferometer, together with a photon detection  during the
spin-measuremement/preparation cycle. To verify heralded quantum
teleportation, we focus on photons emitted in an $800$ps long
time-interval (as depicted in Figs.~3a), which is slightly longer
than the QD lifetime (650ps).


In order to demonstrate classical  correlations between the color of
the photon to be teleported and the final spin state, we use an
input photon that is prepared either in $|\omega_r\rangle_A$ or
$|\omega_b\rangle_A$. For a mode A photon in $|\omega_r\rangle_A$, a
three-fold coincidence projects the photon in mode B onto
$|\omega_b\rangle_B$ and the spin onto $|\uparrow\rangle$. Figure~3b
shows that the same-period ($period$ $0$) 3-fold coincidences
corresponding to a spin measurement in $|\uparrow\rangle$ are a
factor $\sim 4$ larger than those corresponding to
$|\downarrow\rangle$. By using an input photon in state
$|\omega_b\rangle_A$, the $period$ $0$ 3-fold coincidences show that
detecting the spin in $|\downarrow\rangle$ is now $\sim 4.0$ more
likely than detecting it in $|\uparrow\rangle$ (Fig.~3c), in
accordance with the predictions of the teleportation protocol. From
these measurements, we obtain the teleported state fidelities
$0.79\pm0.1$ ($0.82\pm0.09$) for $|\omega_r\rangle_A$
($|\omega_b\rangle_A$). We note that the time-resolved three-fold
coincidence counts for the input state $\left| \omega_b
\right\rangle_A$ (Fig.~3d) show oscillations at $\omega_b - \omega_r
= 4.9$GHz.

Demonstration of quantum teleportation requires that coherences in
the  photonic superposition state at the input mode A are faithfully
transferred onto the spin state. To verify this, we prepare the
single-photon in mode A in either $|\omega_r\rangle_A +
|\omega_b\rangle_A$ or $|\omega_r\rangle_A - |\omega_b\rangle_A$.
Since the propagation time of the photons onto the superconducting
single-photon detectors (SSPD) is about $11$ns, we introduce a
spin-echo pulse sequence to ensure that the spin measurement is
carried out only after the coincidence detection at the output of
the HOM interferometer. The 3-fold coincidences now indicate an
enhanced probability for detection of the spin in state
$|\uparrow\rangle + |\downarrow\rangle$ for an input photon in
$|\omega_r\rangle_A + |\omega_b\rangle_A$ and $|\uparrow\rangle -
|\downarrow\rangle$ for an input photon in $|\omega_r\rangle_A -
|\omega_b\rangle_A$ (Figs.~3e,f,g). From these measurements, we
obtain the teleported state fidelities $0.76\pm0.03$ ($0.75\pm0.03$)
for $|\omega_r\rangle_A + |\omega_b\rangle_A$($|\omega_r\rangle_A -
|\omega_b\rangle_A$). The overall teleportation fidelity  for the
above 4 different input photonic states is $F_T = 0.78 \pm 0.03$,
which is above the classical threshold of $2/3$ by 3.7 standard
deviations.

The measured teleportation fidelity is limited by the small mismatch
between the temporal pulse-shapes and the spatial overlap profiles
of the two interfering photons, as well as the finite spin-photon
entanglement fidelity. We emphasize that, unlike measurements of
entanglement fidelity \cite{Gao12,DeGreve12,Schaibley13}, the
experimentally determined teleportation fidelity is independent of
the detector jitter. Prolongation of the spin-echo time in our
experiments is limited by dynamical nuclear spin polarization
effects \cite{Press10,Urbazsek13} that are omnipresent in
self-assembled QDs (Supplementary Information).

We expect quantum teleportation from a propagating  photonic qubit
to a stationary solid-state spin qubit to play a central role in
quantum networks where nodes incorporating a small number of
spin-qubits are interconnected using photons. While the finite
efficiency of quantum state transfer is a limitation in general, the
heralding of successful transfer renders the implemented protocol
relevant for schemes such as linear optics quantum computation which
rely on post-selection on measurement results and where quantum
memory for the photons is crucial. Since our photonic qubit is
generated by a neutral QD exciton, our experiment can also be
considered as the realization of teleportation from the exciton
qubit of one QD to the spin qubit of another QD. A natural extension
of the quantum teleportation protocol is to replace the photonic
qubit with a photon that is entangled with another QD spin;
coincidence detection at the output of the HOM interferometer in
this case will herald successful generation of a maximally entangled
state of two distant spins that have never directly interacted with
each other \cite{Moehring07,Bernien13}. Further interesting
extensions include quantum teleportation to a decoherence-avoiding
singlet-triplet qubit in a QD molecule \cite{Weiss12},  to a
hole-spin \cite{Brunner09}, or a subsequent transfer of the spin
state from the self-assembled QD to a gate-defined QD, where longer
coherence times on the order of $200 \mu$s and coherent two-qubit
gates have been demonstrated \cite{Bluhm10}.

{}

\newpage
\begin{figure}
\includegraphics[scale=1.0]{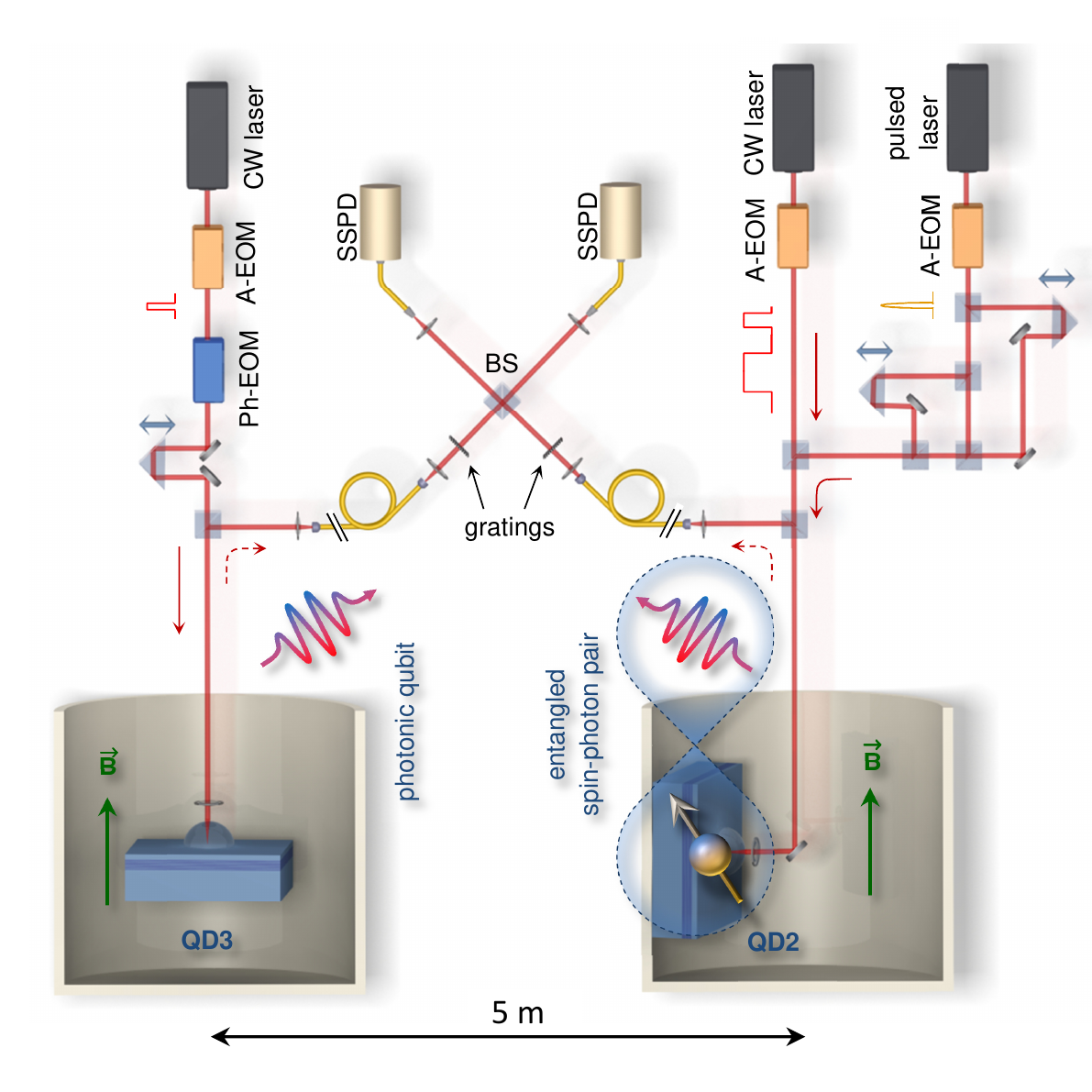}
\caption{The neutral exciton transitions in
a single InGaAs quantum dot (QD3) are resonantly excited by a $0.4
ns$ laser pulse. Upon spontaneous emission the QD generates a photon
in an experimentally tunable superposition state of the two neutral
exciton transition frequencies, forming a single photon qubit. A
different quantum dot (QD2) in the single-electron charged state,
under a perpendicular external magnetic field, is resonantly excited
by a sequence of pulses that initialize the electron spin and excite
the blue negatively charged exciton (trion). An entangled
spin-photon pair is generated upon spontaneous emission from the
trion state. The two QDs are located in separate liquid Helium bath
cryostats that are $\sim 5 m$ apart.  
The emitted single photons from
the two QDs are collected into single mode fibers and sent to
interfere on a beam splitter in a Hong-Ou-Mandel setup. In both
arms, transmission gratings with $70GHz$ bandwidth are used to
filter out the $200GHz$ detuned $4ps$ rotation laser as well as the
phonon sideband emission from the QDs. Superconducting single photon
detectors (SSPD) are used to detect the photons on the output arms
of the beam splitter. A coincidence detection on the two SSPDs
heralds a successful teleportation of the state of the photonic
qubit to that of the electron spin in QD2, which is measured by the
same two SSPDs at a later time.}
\end{figure}

\clearpage

\begin{figure}
\includegraphics[scale=0.9]{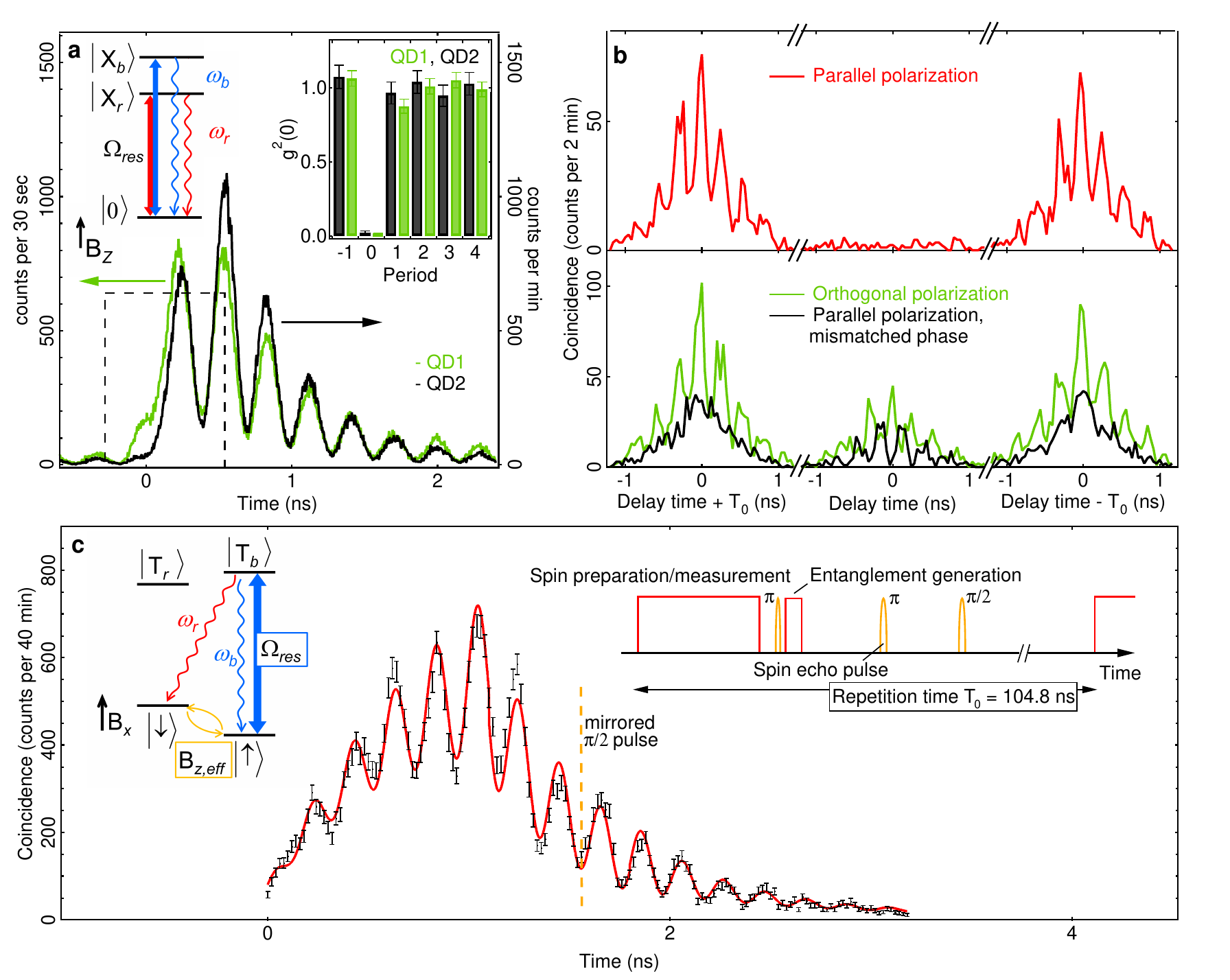}
\end{figure}

\clearpage

{\bf Fig.2: Characterization of the photonic qubit, photon
indistinguishability  and entanglement generation.} ({\bf a}) Time
resolved resonance fluorescence counts are plotted for QD1 (green)
and QD2 (black). A $0.8 ns$ two color laser pulse (dashed line) that
is resonant with both excitonic transitions $|X_b\rangle$ and
$|X_r\rangle$ excites the QD (upper left panel); subsequent
spontaneous emission generates a single photon in a superposition of
$\omega_b$ and $\omega_r$. The matched oscillations of the emission
from the two QDs at $\Delta = \omega_b - \omega_r = 3.45 GHz$
indicate that the generated photonic qubit states are nearly
identical. The upper right panel shows photon correlation $(g^2)$
measurements on the two QDs using the emitted photons detected after
the excitation pulse is turned off, indicating that the likelihood
of two or more photon detection events is very small. ({\bf b}) The photonic qubits depicted in {\bf a} are incident on a
beam splitter. Coincidence counts on the two output arms of the beam
splitter are plotted as a function of the delay between the recorded
photon arrival times. $T_0$ is the pulse repetition time of $13.1
ns$. As the photons are indistinguishable when the input modes have
parallel polarization (top panel), the coincidence counts within the
time window [-1.2 ns, 1.2 ns] are 11 times smaller than those in
which the second photon is detected in other periods. In the bottom
panel, the photons are rendered distinguishable either through their
polarization (green) or their phase (black), leading to a center
peak coincidence count that is 2 times smaller than the other
periods. ({\bf c}) QD2 is prepared in the single-electron charged state.  An
external magnetic field of $B_x = 0.7 T$ is applied in the Voigt
geometry leading to the energy-level diagram depicted in the left
panel. The pulse sequence used for the generation of spin-photon
entangled state and measurement of correlations after a spin-echo
sequence is shown in the upper right panel. The center panel shows
the photons emitted during and after the entanglement generation
pulse, conditioned upon a spin measurement in the
$\frac{1}{\sqrt{2}}(|\uparrow \rangle - |\downarrow \rangle)$ state.
The oscillations in the depicted coincidence counts at $\Delta =
\omega_b - \omega_r = 4.9 GHz$ are due to the beating between the
two frequency components of the projected photonic superposition
state.

\begin{figure}
\includegraphics[scale=1.3]{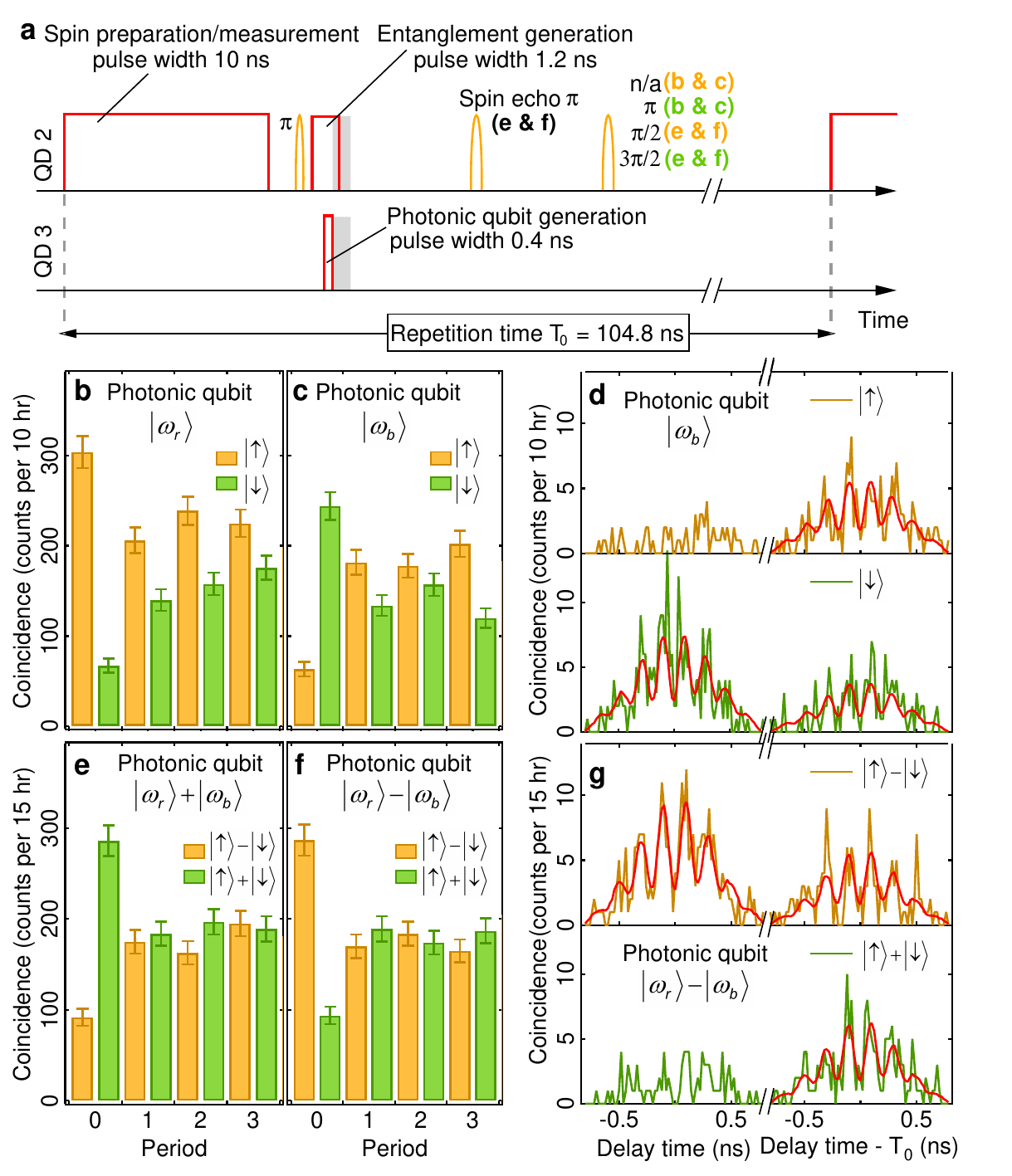}
\end{figure}
\clearpage

{\bf Fig. 3: Demonstration of quantum teleportation.} ({\bf a})  The
teleportation pulse sequence applied to QD2 for entanglement
generation and spin measurement, as well as to QD3 for photonic
qubit generation (see Fig. 2 for details). ({\bf b, c}) Classical
correlations between the state of the photonic qubit and the state
of the electron spin: the plots show $3$-fold coincidence counts
between the two output arms of the beam splitter during the $0.8 ns$
time interval depicted by the grey box in {\bf a} and a photon
detection during the following spin measurement pulse $(period = 0)$
or during a later pulse period $(period > 0)$. An enhanced
probability for the spin state $|\uparrow \rangle$ ($|\downarrow
\rangle$) as well as a decreased probability for $|\downarrow
\rangle$ ($|\uparrow \rangle$) is observed when the photonic qubit
was initially prepared in $|\omega_r\rangle$ ($|\omega_b\rangle$).
The coincidence counts, measured for $period > 0$, differ for
measurement of the spin in $|\uparrow \rangle$ and $|\downarrow
\rangle$ states, due to the fact that the unconditioned $|\uparrow
\rangle$ population is $>  0.5$; this is a consequence of excitation
probability to the $|T_b\rangle$ state during the entanglement
generation pulse being $< 1$. ({\bf d}) Time resolved $3$-fold
coincidence counts corresponding to the data plotted in the first
two periods in {\bf c}. Red lines are guides to the eye, showing the
expected oscillations for coincidences. ({\bf e, f}) Quantum
correlations between the state of the photonic qubit, prepared
either in $|\omega_r\rangle + |\omega_b\rangle$ {\bf e} or
$|\omega_r\rangle - |\omega_b\rangle$ {\bf f}, and the spin qubit.
The $3$-fold coincidence counts are measured as described above. A
spin echo $\pi$-pulse is applied to prolong the spin coherence time.
Detection of a coincidence on the beam splitter in {\bf e} ({\bf f})
projects the spin into $|\uparrow \rangle + |\downarrow \rangle$
($|\uparrow \rangle - |\downarrow \rangle$), as evidenced by the
observed correlations. ({\bf g}) Time resolved coincidence counts
corresponding to the data plotted for the first two periods in {\bf
f}. Red lines are guides to the eye.

\end{document}